\documentclass[twocolumn]{aastex62}

\usepackage{booktabs} 
\usepackage{multirow} 
\usepackage{makecell} 
\usepackage{ragged2e} 
\usepackage{amsmath} 
\usepackage{comment}
\usepackage{xcolor} 
\usepackage{hyperref}

\newcommand{\Mj}{$M_{\rm Jup}$}

\newcommand{\kms}{km\,s$^{-1}$}

\newcommand{\twCO}{$^{12}$CO}
\newcommand{\thCO}{$^{13}$CO}
\newcommand{\thCOfull}{$^{13}$CO\,$J=2\rightarrow1$}

\newcommand{\CtwoH}{C$_{\rm 2}$H}
\newcommand{\CtwoHfull}{C$_{\rm 2}$H $N=3\rightarrow2$ $J=5/2\rightarrow3/2$}
\newcommand{\hcnfull}{HCN $J=3\rightarrow2$}

\newcommand{\discminer}{\textsc{discminer}}
\newcommand{\disksurf}{\textsc{disksurf}}

\newcommand{\hd}{HD\,163296}

\graphicspath{{./}{figures/}}

\received{September 2025} 
\accepted{December 2025}

\shorttitle{CPD candidate in the disk of \hd{}}
\shortauthors{Izquierdo et al.}

\begin{document}


\title{%
  \begin{minipage}{0.95\textwidth}
    \centering
    \large\bfseries
    Circumplanetary Disk Candidate in the Disk of HD\,163296 Traced by Localized Emission from Simple Organics
  \end{minipage}
}


\author[0000-0001-8446-3026]{Andr\'es F. Izquierdo}
\altaffiliation{NASA Hubble Fellowship Program Sagan Fellow}
\affiliation{Department of Astronomy, University of Florida, Gainesville, FL 32611, USA; \textrm{\url{andres.izquierdo.c@gmail.com}}}

\author[0000-0001-7258-770X]{Jaehan Bae}
\affiliation{Department of Astronomy, University of Florida, Gainesville, FL 32611, USA; \textrm{\url{andres.izquierdo.c@gmail.com}}}

\author[0000-0002-5503-5476]{Maria Galloway-Sprietsma}
\affiliation{Department of Astronomy, University of Florida, Gainesville, FL 32611, USA; \textrm{\url{andres.izquierdo.c@gmail.com}}}

\author[0000-0001-7591-1907]{Ewine F. van Dishoeck}
\affiliation{Leiden Observatory, Leiden University, 2300 RA Leiden, The Netherlands}
\affiliation{Max-Planck Institut für Extraterrestrische Physik (MPE), Gießenbachstr. 1 , 85748, Garching bei München, Germany}

\author[0000-0003-4689-2684]{Stefano Facchini}
\affiliation{Dipartimento di Fisica, Universit\`a degli Studi di Milano, Via Celoria 16, 20133 Milano, Italy}

\author[0000-0003-4853-5736]{Giovanni Rosotti} 
\affiliation{Dipartimento di Fisica, Universit\`a degli Studi di Milano, Via Celoria 16, 20133 Milano, Italy}

\author[0000-0002-0491-143X]{Jochen Stadler} 
\affiliation{Universit\'e C\^ote d'Azur, Observatoire de la C\^ote d'Azur, CNRS, Laboratoire Lagrange, 06304 Nice, France}

\author[0000-0002-7695-7605]{Myriam Benisty}
\affiliation{Max-Planck Institute for Astronomy (MPIA), Königstuhl 17, 69117 Heidelberg, Germany}

\author[0000-0003-1859-3070]{Leonardo Testi}
\affiliation{Dipartimento di Fisica e Astronomia, Università di Bologna, I-40129 Bologna, Italy}

\begin{abstract}

Atacama Large Millimeter/submillimeter Array observations suggest that the disc of \hd{} is being actively shaped by embedded, yet unseen protoplanets, as indicated by numerous gas and dust substructures consistent with planet-disc interaction models. We report the first detection of simple organic molecules, HCN and \CtwoH{}, tracing a candidate circumplanetary disc (CPD) in the \hd{} system, located at an orbital radius of $R=88\pm7$\,au and azimuth $\phi=46\pm3^\circ$ (or $R=0\farcs{75}$, $\rm{{PA}}=350^\circ$ in projected sky coordinates), and originating near the midplane of the circumstellar disc. The signature is localised but spectrally resolved, and it overlaps with a previously reported planet candidate, P94, identified through kinematic perturbations traced by CO lines. We propose a scenario in which the observed chemical anomalies arise from increased heating driven by the forming planet and ongoing accretion through its CPD, facilitating the thermal desorption of species that would otherwise remain frozen out in the disc midplane, and potentially triggering the activation barriers of chemical reactions that lead to enhanced molecular production. Based on a first-order dynamical analysis of the HCN spectrum from the CPD---isolated with a 7$\sigma$ significance---we infer an upper limit on the planet mass of 1.8\,\Mj{}, consistent with predictions from CO kinematics and constraints from direct imaging studies. By comparing the CPD sizes derived from our models with theoretical expectations where the CPD radius corresponds to roughly one-third of the planet’s Hill radius, we favor CPD gas temperatures $T > 150$\,K, planet masses $M_{\rm p} < 1.0$\,\Mj{}, and CPD radii $R_{\rm CPD} < 2$\,au.
\end{abstract}


\section{Introduction} \label{sec:intro}

\begin{figure*}
   \centering
   \includegraphics[width=1.0\textwidth]{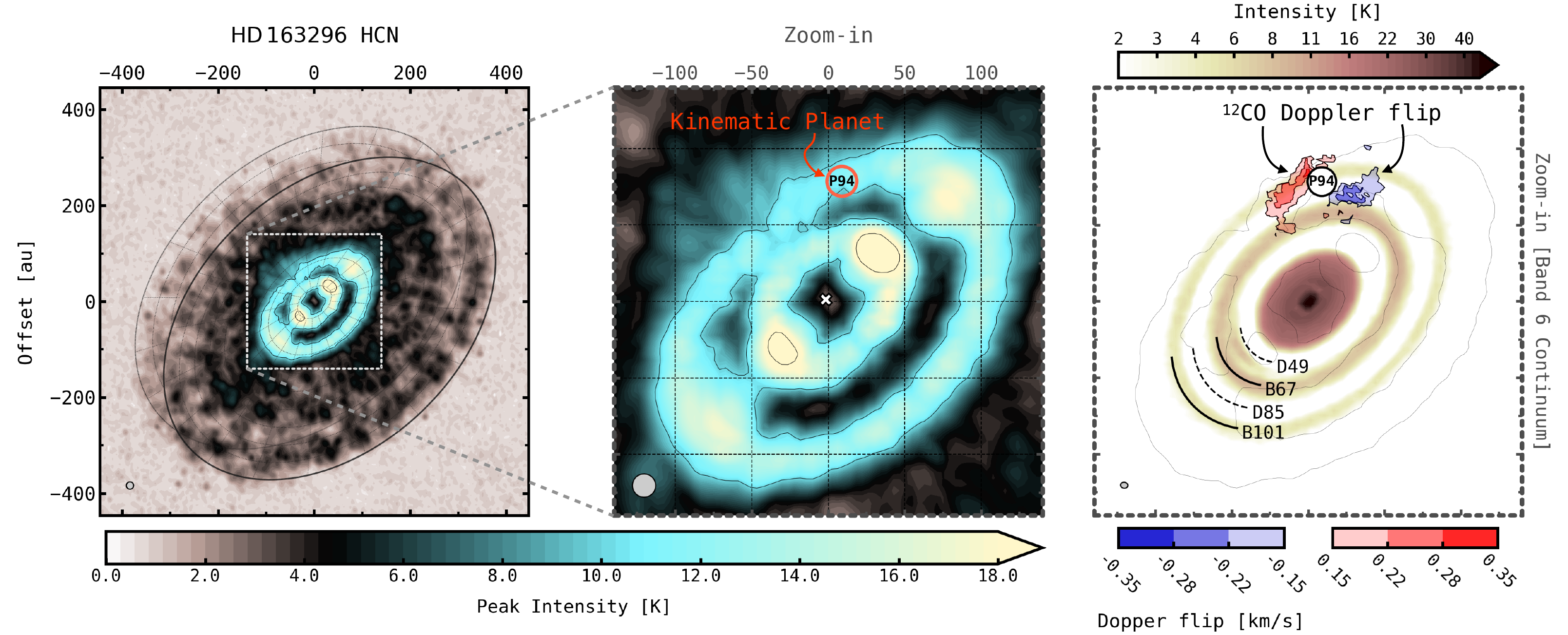} 
      \caption{HCN and millimeter dust continuum emission from the disc of \hd{} as observed with ALMA. The left panel shows the absolute peak intensity of \hcnfull{} line data from the MAPS project \citep{oberg+2021}, computed using the Rayleigh-Jeans approximation, while the middle panel provides a zoomed-in view of the central region. For the same section, the right panel displays the millimeter dust continuum emission from DSHARP \citep{andrews+2018, isella+2018}, with dashed and solid lines marking the radial locations of selected continuum gaps (DXX) and rings (BXX), respectively. Overlaid is the localised \twCO{} velocity perturbation identified by \citet{izquierdo+2022}, along with contours of HCN emission at $T=10$\,K for reference. The circle labeled P94 marks the location of the planet candidate associated with this velocity feature, which we propose is also responsible for the localised intensity signals observed in the channel maps of HCN and \CtwoH{} lines (see Sect. \ref{subsec:signals}).
              }
         \label{fig:peak+continuum}
\end{figure*} 


The disc of \hd{} has been identified as a promising host of ongoing planet formation, as evidenced by multiple annular signatures observed in both dust continuum and molecular line emission with the Atacama Large Millimeter/submillimeter Array \citep[ALMA;][]{degregorio+2013,isella+2016,isella+2018}. Kinematic studies, also enabled by ALMA, further support a planetary origin for these features. Concentric deviations from Keplerian rotation around the dust substructures can be explained by invoking gap-carving, Jupiter-mass planets within the inner $R<160$\,au of the disc \citep{teague+2018a}, while in the outer regions large-scale velocity perturbations point to the possibility of an additional giant planet on a wide orbit \citep[$R\sim260$\,au,][]{pinte+2018b, calcino+2022}. 

More recently, \citet{izquierdo+2022, izquierdo+2023} further investigated the kinematics of this system, detecting a strongly localised velocity perturbation traced by \twCO{} and \thCOfull{} emission lines. Based on empirical comparisons with hydrodynamical models, the authors attributed this feature to a 1\,\Mj{} protoplanet (P94), located at the centre of the perturbation, at an orbital radius of $R=94\pm6$\,au and an azimuth of $\phi=50\pm3^\circ$ in disc-frame coordinates, corresponding to a projected separation of $R=0\farcs{77}$ from the disc centre and a position angle of $\rm{{PA}}=352.1^\circ$, measured from north to east. The inferred mass of P94 is consistent with the nondetections at near-infrared wavelengths with VLT/SPHERE \citep{mesa+2019} and JWST/NIRCam \citep{uyama+2025}, which suggest upper limits of $\sim$4\,\Mj{} at this orbital separation, but remains sufficient to carve a dust and gas gap in the disc \citep{teague+2018a}.

Nevertheless, although crucial for indirect detection, CO lines are known to originate from intermediate to high altitudes above the midplane of this disc \citep[$z/r \approx 0.15$–$0.3$;][]{law+2021_maps4, paneque+2023, izquierdo+2023}, making the derived planet mass model-dependent---sensitive to assumptions about disc properties such as viscosity and vertical pressure structure \citep{rabago+2021}---and more susceptible to the influence of hydrodynamical instabilities than the midplane layers \citep{barraza-alfaro+2024}. 

Luckily, deep ALMA observations offer the opportunity to broaden our understanding of planet-driven signatures by probing a variety of molecular species that trace distinct radial and vertical regions of the disc, depending on their abundances and excitation conditions \citep[see reviews by][]{miotello+2023, oberg+2023}. Of particular interest for planet searches are tracers of outflows and shock chemistry---such as SO, SiS, and SiO---some of which have recently been detected in discs near previously reported planet candidates \citep{booth+2023,law+23_2023ApJ...952L..19L}, as well as in discs with strong dynamical perturbations potentially driven by planets \citep{zagaria+2025}. Another important class of tracers accessible with ALMA includes molecules like HCN and \CtwoH{}, whose abundances are highly sensitive to high-temperature chemistry \citep{boonman+2001, doty+2002, agundez+2008}, the strength of the ultraviolet (UV) radiation field \citep{agundez+2018, visser+2018}, and the local C/O ratio \citep{cleeves+2018}. Consequently, thermochemical models predict that the gas-phase distribution of these molecules may be enhanced in the vicinity of giant planets owing to increased heating and UV irradiation associated with ongoing accretion onto the forming planet \citep{cleeves+2015,aoyama+2020,calahan+2023}

In this Letter, we present the first evidence for the presence of these simple organics in a circumplanetary region still embedded within its parent circumstellar disc (CSD). The system orbits the Herbig Ae star \hd{}, located at a distance of 101.5\,pc from Earth at ICRS equatorial coordinates $\alpha = 17^{\mathrm{h}}56^{\mathrm{m}}21\fs$, $\delta = -21\degr57\arcmin21\arcsec$ \citep{bailer-jones+2018}. The detected signatures manifest as localised intensity enhancements in HCN and \CtwoH{} lines, overlapping with the \twCO{} and \thCO{} velocity perturbations previously attributed to the kinematic planet candidate P94. Furthermore, these molecular anomalies appear to arise near the CSD midplane, suggesting an origin in the immediate vicinity of P94 and possibly tracing emission from a circumplanetary disc (CPD) around the planet.

CPDs are natural by-products of the planet formation process \citep{miki+1982,tanigawa+2002}, yet observational evidence for them remains scarce, with detections reported in only two systems where the circumstellar disc material is already significantly depleted \citep{isella+2019, benisty+2021, bae+2022}. As a result, knowledge of key CPD properties such as temperature, size, and dynamical mass is also limited. In this work, we provide estimates of these parameters for the CPD candidate around P94 through a first-order analysis of the HCN line profile observed in its vicinity. These properties represent valuable proxies for the accretion dynamics and evolutionary stage of the protoplanet \citep{szulagyi+2016, szulagyi_mordasini+2017, fung+2019}, and are crucial for understanding the onset of satellite formation in the system \citep{canup+2002}.

\section{Observations and disc modelling}

\subsection{Datasets}

The core of this work is based on archival ALMA data (Project ID: 2018.1.01055.L) targeting HCN in its rotational transition $J=3\rightarrow2$, which is split into multiple hyperfine components. These data were originally reduced and analysed by the MAPS collaboration \citep{oberg+2021}, which provides three fiducial, continuum-subtracted position-position-velocity cubes for this tracer, referred to as HF2, HF1, and HF3 (in order of ascending frequency), each centred on a different hyperfine transition within the group. HF2 and HF3 are centred on the faint peripheral $F=3\rightarrow3$ and $F=2\rightarrow2$ components, respectively, while HF1 is centred on the bright $F=3\rightarrow2$ transition, at a rest frequency of $\nu=265.886499$\,GHz \citep{muller+2001}. At the channel spacing of the data, the $F=3\rightarrow2$ line is blended with other bright components, $F=2\rightarrow1$ and $F=4\rightarrow3$. Under the assumption of local thermodynamic equilibrium, these lines account for $\sim92\%$ of emission in the $J=3\rightarrow2$ group \citep[e.g.][]{mullins+2016}, and thus our analysis is dominated by their combined signal. 

For the modelling of the HCN channel maps from the circumstellar disc and the candidate CPD we use the JvM-corrected HF3 cube, centred on the highest-frequency component at  $\nu=265.888522$\,GHz, in which the localised intensity signal around the planet P94 was first identified. This choice, however, has no practical impact on the analysis, since all bright hyperfine components in the group are closely spaced in frequency and fall within the spectral coverage of any of the three fiducial cubes. Accordingly, the P94 intensity signal is also detected in the HF1 and HF2 cubes, both with and without JvM correction, as illustrated in Appendix \ref{sec:figures}. We also use a fiducial datacube of \CtwoH{} emission provided by MAPS, focusing on the brightest transition, $F=4\rightarrow3$ at $\nu=262.004226$\,GHz, in the hyperfine group $N=3\rightarrow2$, $J=7/2\rightarrow5/2$, where we identify spatial overlap with the localised HCN signal around P94. The synthesized beam size of all datasets used in this work is $0\farcs{15} \times 0\farcs{15}$, and the velocity channels are spaced by $0.2$\,\kms{}, with rms noise levels of 0.43\,mJy\,beam$^{-1}$ for HCN and 0.53\,mJy\,beam$^{-1}$ for \CtwoH{}. Details of the data calibration and imaging procedures are provided in \citet{czekala+2021}.

\begin{figure*}
   \centering
   \includegraphics[width=1.0\textwidth]{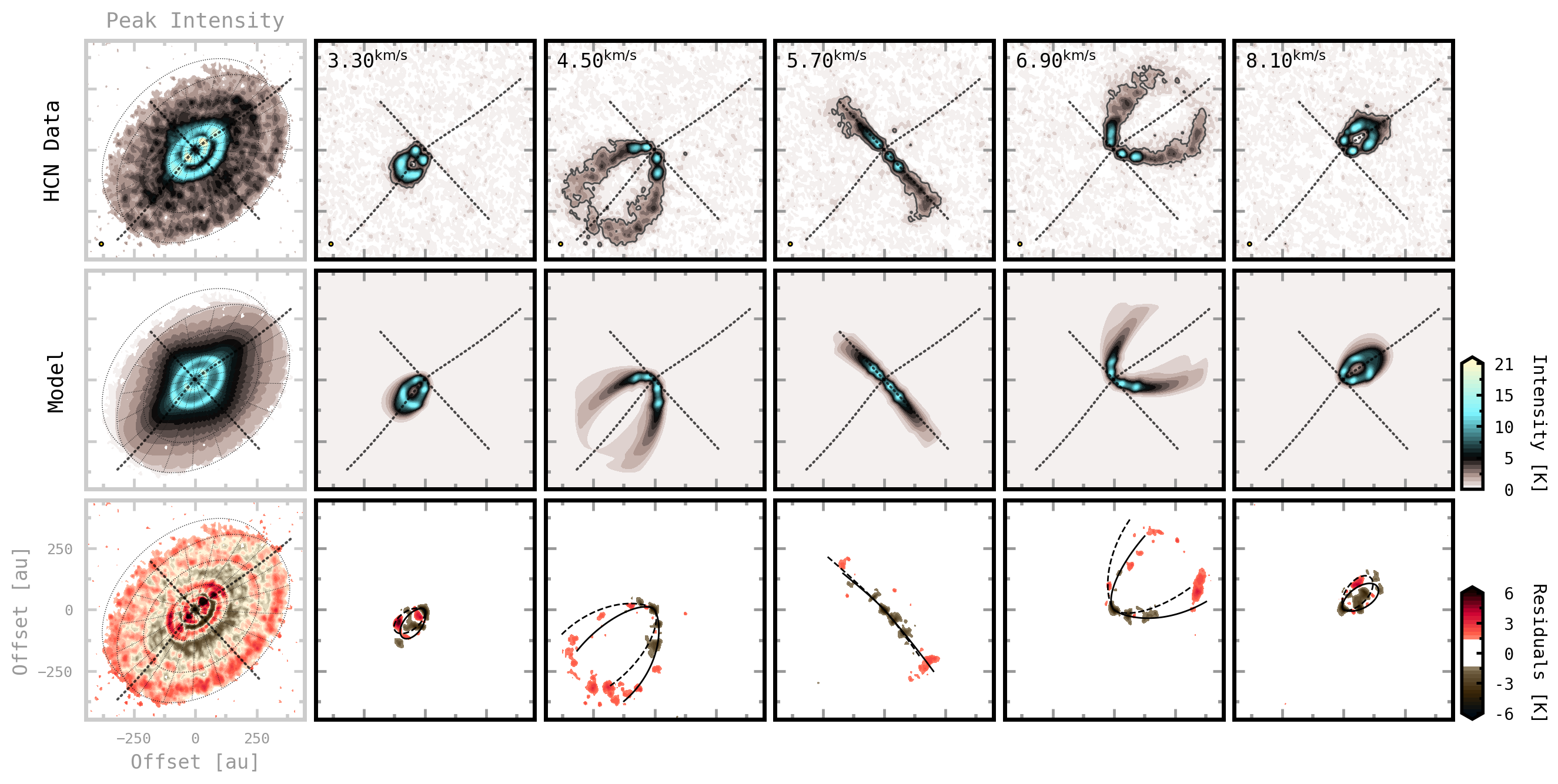} 
      \caption{Selected channel maps of HCN line intensity from the \hd{} disc (top row), compared with those from the best-fit \discminer{} model (middle row), computed using the Rayleigh-Jeans approximation. Contours in the top row enclose emission above five times the rms noise level of 0.33\,K. Also shown are intensity residuals for each velocity channel (bottom row), with overlaid isovelocity contours from the model upper and lower surfaces as solid and dashed lines, respectively. For reference, the best-fit systemic velocity is $\upsilon_{\rm sys}=5.79$\,km\,s$^{-1}$. The synthesized beam of the observations is shown in the bottom left corner of the panels in the top row. Residuals with amplitudes lower than four times the rms noise have been masked.
              }
         \label{fig:channels}
\end{figure*} 


\subsection{Model of the Disc HCN Intensity and Dynamics} \label{sec:discminer}

The disc of \hd{} exhibits significant substructure in HCN and \CtwoH{} emission, including annular enhancements at 48\,au and 114\,au, a prominent dip at 79\,au, and a central cavity in both cases \citep{law+2021_maps3}, consistent with the similar formation pathways of these molecules \citep{guzman+2021}. Figure \ref{fig:peak+continuum} illustrates these annular features for \hcnfull{}, showing the absolute peak intensity map of the line alongside the Band 6 dust continuum, overlaid with the \twCO{} Doppler-flip signature associated with the kinematic planet P94 \citep{izquierdo+2022}. We note that, since the HCN peak intensity is dominated by emission from the circumstellar disc, the localised HCN signal associated with P94 is not visible in this map. Instead, it becomes apparent when the channel maps of the tracer are examined individually, as discussed in Sect.~\ref{subsec:signals}.  

To model the channel-by-channel intensity and kinematics of the disc while capturing its radial substructures, we first ran \discminer{}\footnote{\url{https://github.com/andizq/discminer}} to retrieve baseline geometric and line-profile parameters assuming radially decreasing peak intensity and line width \citep[see][for details]{izquierdo+2025}, with the disc inclination fixed to the best-fit continuum value from \citet{huang+2018_incl}. In a second step, we fitted a customised version of the radial peak intensity, including two parametric Gaussian rings, while keeping all other parameters fixed. We restricted the modelling to the HCN data to avoid contamination from the strong hyperfine components present in \CtwoH{}. Table \ref{table:attributes_parameters} summarises the functional forms of the adopted model and the resulting best-fit parameters, and Figure \ref{fig:channels} presents selected HCN intensity channel maps, together with the corresponding best-fit model and residuals.

\setlength{\tabcolsep}{6pt} 

\begin{table*}
\centering
{\renewcommand{\arraystretch}{1.5}
 \caption{List of attributes adopted in our \discminer{} channel-map model of the \hcnfull{} emission line from the disc of \hd{}, along with the corresponding best-fit parameters.}
  \label{table:attributes_parameters}
\begin{tabular}{ l|l|l } 

\toprule
\toprule
Attribute & Prescription &  \multicolumn{1}{c}{Best-fit parameters for \hcnfull{} } \\
\hline

Orientation & uniform $i$, PA, $x_c$, $y_c$ &  $i=46.7^\circ$ [Fixed] \quad $\rm{PA} = 312.8^\circ$ \quad $x_c=-1.5$\,mas \quad $y_c=32.0$\,mas \\

\hline

Velocity & $\upsilon_{\rm k} = \sqrt{\frac{GM_\star}{r^3}}R$, $\upsilon_{\rm LSRK}$  & $M_\star=1.99$\,M$_\odot$ \quad $\upsilon_{\rm LSRK}=5.79$\,km\,s$^{-1}$  \\

\hline

Upper surface & $z = z_{0} (R/D_0)^p \exp{[-(R/R_t)^q]}$ & $z_0 = 21.7$\,au \quad $p=1.50$ \quad $R_t = 429.7$\,au \quad $q=1.03$ \\

Lower surface & $z = -z_{0} (R/D_0)^p \exp{[-(R/R_t)^q]}$ & $z_0 = 17.3$\,au \quad $p=1.44$ \quad $R_t = 456.5$\,au \quad $q=0.77$ \\

\hline
Peak intensity & $I_p (R \leq R_{\rm out})$ = $I_0 (R/D_0)^p (z/D_0)^q$ & $I_0 = 124.4$\,mJy\,pix$^{-1}$ \quad $p=-4.31$ \quad $q=3.04$ \quad $R_{\rm out}=405.7$\,au \\

Line width & $L_w = L_{w0} (R/D_0)^p (z/D_0)^q$ & $L_{w0} = 0.07$\,km\,s$^{-1}$ \quad $p=0.42$ \quad $q=-1.07$  \\ 

Line slope & $L_s = L_{s0} (R/D_0)^p$ & $L_{s0} = 1.42$ \quad $p=0.24$  \\

\hline

Intensity ring \#1 & $I_1 = B_1 \exp\left(-\dfrac{(R - \mu_1)^2}{2 \sigma_1^2}\right) $ & $B_1 = 1.5$\,mJy\,pix$^{-1}$ \quad $\mu_1=47.6$\,au \quad $\sigma_1=2.5$\,au \\

Intensity ring \#2 & $I_2 = B_2 \exp\left(-\dfrac{(R - \mu_2)^2}{2 \sigma_2^2}\right) $ & $B_2 = 0.7$\,mJy\,pix$^{-1}$ \quad $\mu_2=114.3$\,au \quad $\sigma_1=7.3$\,au \\

\bottomrule

\end{tabular}


\justifying
{\noindent \textbf{Note.} $G$ is the gravitational constant. $D_0=100$\,au is a normalisation factor. In disc coordinates, $z$ denotes the emission height above the midplane, $R$ the cylindrical radius, and $r$ the spherical radius. The composite model peak intensity is given by $I_f = I_p + I_1 + I_2$, and is set to zero for radii beyond $R_{\rm out}$. The systemic velocity, $\upsilon_{\rm LSRK}$, was shifted by 2.36\,km\,s$^{-1}$ to align with the rest frequency of the reference $J=3\rightarrow2$, $F=3\rightarrow2$ transition. 
  }
  
  }
\end{table*}

\section{Results} \label{sec:results}

\subsection{Radial and Vertical Distribution of HCN in the Circumstellar Disc}
\label{subsec:height}

Modelling efforts on the chemical composition of protoplanetary discs suggest that HCN and \CtwoH{} are effective tracers of UV radiation, as they can be efficiently synthesized through UV-driven chemistry in warm gas \citep{walsh+2012, agundez+2018, visser+2018}.
These models also indicate that, in the outer and colder regions of discs, HCN is primarily found at intermediate heights ($z/R \approx 0.1-0.3$) above the midplane, where it likely forms via gas-phase chemistry involving ion-molecule reactions followed by dissociative recombination. More generally, the precise radial and vertical distributions of HCN and other such molecules sensitive to the UV radiation field depend strongly on the local UV opacity influenced by dust grain growth, vertical settling, and radial drift, which control how deeply UV photons can penetrate into the gas disc \citep{jonkheid+2007, henning+2010, cazzoletti+2018, calahan+2023}. While the primary goal of this Letter is to report the detection of planet-driven HCN and \CtwoH{} intensity signatures in \hd{}, it is first necessary to characterise the radial and vertical emission structure of the host circumstellar disc (CSD) in order to provide a robust estimate of the three-dimensional location of these signals relative to the background substructures.

\begin{figure*}
   \centering
   \includegraphics[width=0.9\textwidth]{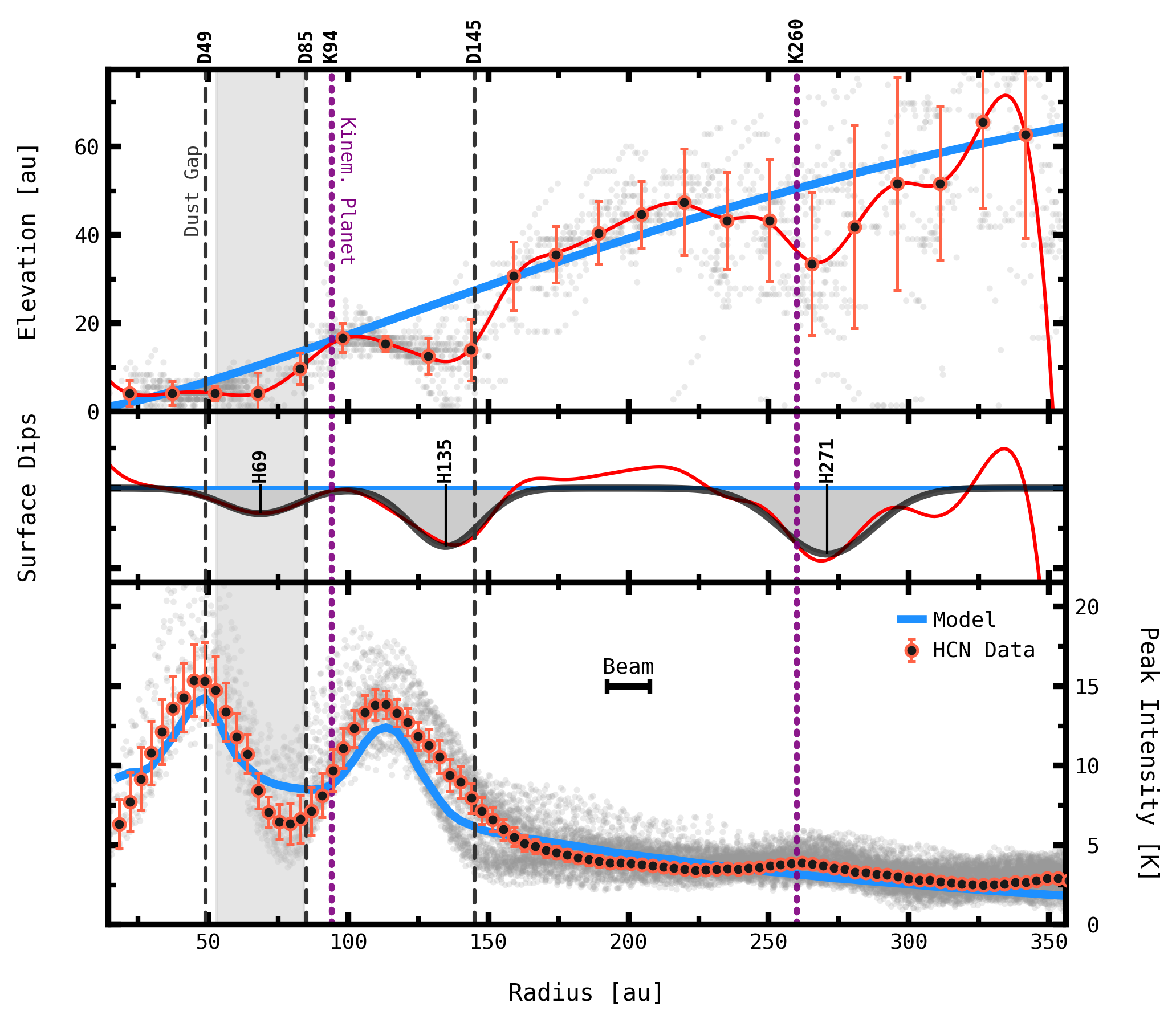}
      \caption{Elevation (top and middle) and peak intensity (bottom) profiles for the front side of the disc of HD\,163296, probed by \hcnfull{} emission, shown for both the best-fit model (blue curves) and the datacube (orange points). The DXX dashed lines mark the locations of the Band 6 dust gaps, while K94 and K260 indicate the orbital separations of previously reported kinematic planet candidates. The surface points in the top panel, extracted with \disksurf{}, reveal significant vertical substructure, including two prominent 30\,au wide dips at 69 and 135\,au, and a tentative dip at 271\,au, labeled as HXX in the middle panel. An optically thin window, highlighted with gray shading and centred on the H69 dip, provides a view of the disc midplane near D85, where a localised signal associated with the P94 planet arises (see Sect. \ref{subsec:signals}). The gray points in the bottom panel are peak intensity values mapped into the disc frame using the model's best-fit geometrical parameters.
              }
         \label{fig:radial_profiles}
\end{figure*}

As illustrated in the top panel of Figure \ref{fig:radial_profiles} our best-fit model is in agreement with a HCN emission surface originating from intermediate layers at $z/R \approx 0.15-0.2$ above the disc midplane, closely matching the emission region of \thCOfull{} \citep{izquierdo+2023}. This contrasts with the other four discs in the MAPS sample, where the bulk of the HCN emission appears to originate from midplane layers ($z/R < 0.1$) despite differences in stellar type and gas temperature \citep[see][for an analysis of the full sample]{paneque+2023}. Such a distinction may indicate lower dust-to-gas mass ratios in those systems, allowing UV photons to penetrate more deeply and thereby push HCN emission to lower altitudes compared to \hd{} \citep{calahan+2023}. We further analysed the HCN emission surface using an independent, non-parametric method with the \textsc{disksurf} package \citep[][]{disksurf, pinte+2018a}, aiming to capture radial modulations in the HCN vertical distribution. This analysis confirms that emission from this tracer arises from layers at $z/R \approx 0.15-0.2$, except for the dips at 67\,au and 135\,au, where the emission surface shifts to altitudes much closer to the midplane ($z/R < 0.1$).

Radially, the HCN intensity profile shows strong modulations within the region where dust substructures are most prominent at millimetre wavelengths ($R<160$\,au), and extends smoothly beyond this area, suggesting that UV-driven or gas-phase chemistry remains active in the cold outer disc \citep{guzman+2021}. To characterise the HCN radial intensity distribution, we fixed all baseline model parameters---derived under the assumption of smooth intensity and Keplerian rotation---and introduced a second modelling step in which two Gaussian components were fitted to better reproduce the channel-by-channel emission of the disc. 

This refined modelling reveals narrow HCN intensity rings centred at 47.6\,au and 114.3\,au, with Gaussian widths (before convolution) of 2.5\,au and 7.3\,au, respectively. At 79\,au lies a prominent intensity dip, with peak brightness temperatures as low as 5\,K, comparable to the low intensities observed in the outer disc (see Fig. \ref{fig:radial_profiles}, bottom panel).
Caution is warranted, however, within the inner $\sim\!40$\,au, where the continuum emission is strong ($>\!20$\,K) and HCN is potentially optically thick \citep{bergner+2021}. In this region, continuum subtraction likely leads to an underestimation of the HCN intensity \citep[see e.g.][]{boehler+2017}, suggesting that the HCN ring at 47.6\,au may in fact extend further inward into the inner disc. Nevertheless, intensity modulations beyond this region appear robust. Using non-continuum-subtracted images we confirm that the subsequent HCN substructures are present at the same locations, with the corresponding dip (81\,au) and peak (103\,au) lying within one beamwidth of the continuum-subtracted values. This is consistent with the findings of \citet{law+2021_maps3}, who reported no clear correlation between chemical and dust substructures across the MAPS sample of discs, suggesting that continuum subtraction is unlikely to play a dominant role in shaping the observed chemical signatures.

Notably, the identified intensity modulations do not consistently correlate with emission surface highs and lows or with the locations of dust rings and gaps, indicating that UV exposure is likely not the sole driver of HCN abundance in the disc; other key factors such as temperature, the local C/O ratio, and the degree of disc flaring may also play a significant role in shaping the molecule's distribution \citep[see][for a discussion]{guzman+2021}.

\subsection{Localised Intensity Signals in HCN and \CtwoH{}} \label{subsec:signals}

\begin{figure}
   \centering
   \includegraphics[width=0.5\textwidth]{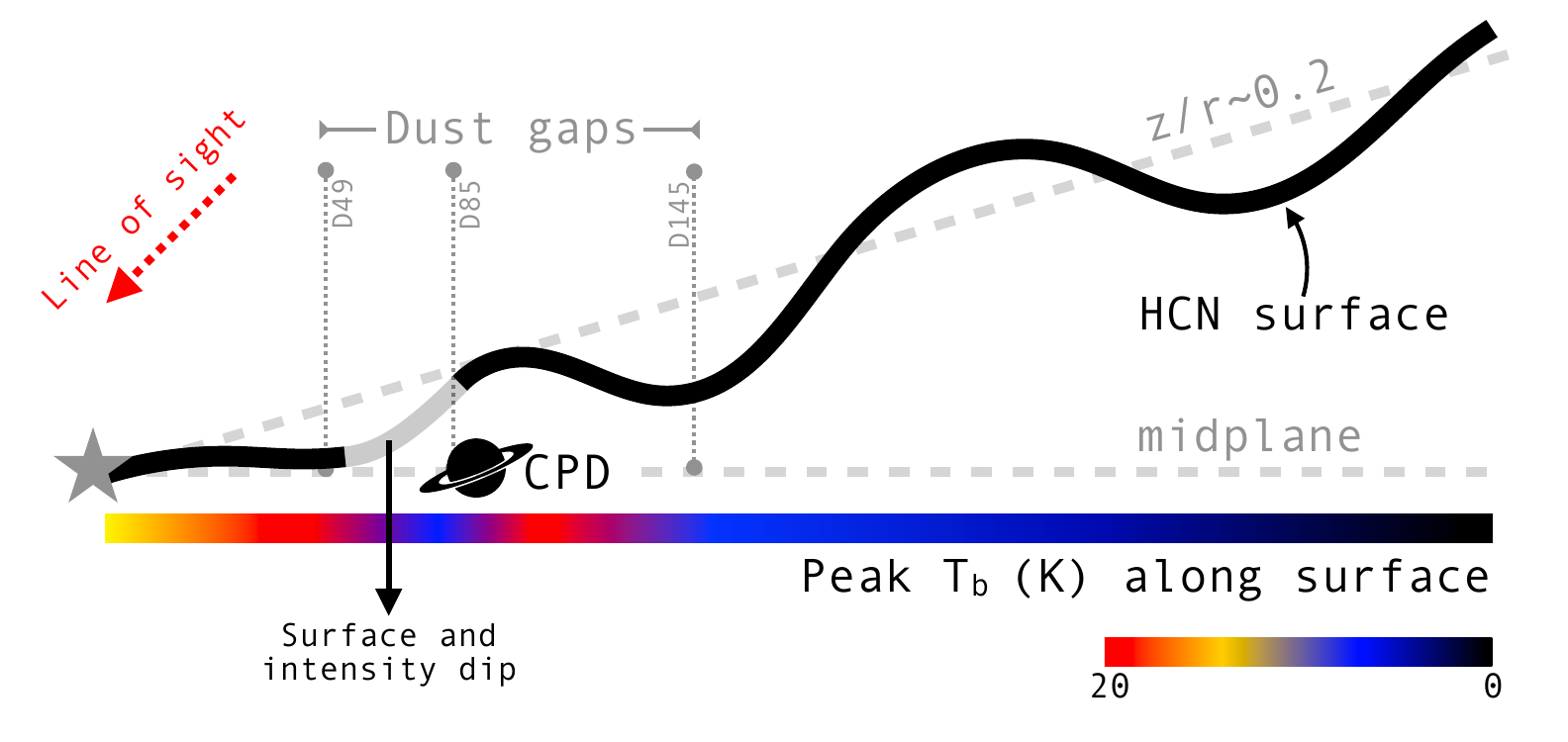}  
      \caption{Cartoon illustrating the radial modulations in the disc surface traced by HCN emission and the relative location of the CPD candidate, whose emission is consistent with an origin near the midplane. This region is partially accessible thanks to a prominent HCN gap near the CPD's orbital separation and the moderate inclination of the disc.
              }
         \label{fig:cartoon}
\end{figure}

A noteworthy feature of the HCN emission from the circumstellar disc is that the dip in surface height between the dust gaps D49 and D85, centred at 69\,au, is also characterized by low optical depths as inferred from hyperfine structure analysis \citep{bergner+2021}. Combined with the system's moderate inclination, this property provides a unique window into regions near the disc midplane and enables a search for chemical signatures originating in the vicinity of any protoplanet potentially associated with the dust gap at 85\,au, as illustrated in Figure \ref{fig:cartoon}. Indeed, within this zone, we detect localised intensity signals\footnote{The outermost regions of the circumstellar disc, beyond $\sim\!140$\,au, are also consistent with optically thin emission, but no localised signals stand out significantly in this area.} that spatially coincide with previously reported \twCO{} and \thCO{} velocity perturbations linked to the Jupiter-mass candidate P94 \citep{izquierdo+2022, izquierdo+2023}. We therefore attribute the origin of these signals to material surrounding the embedded protoplanet, possibly in the form of a circumplanetary disc (CPD). 

The HCN chemical signatures around P94 are highlighted in Figure \ref{fig:channels_mirror}, and in Figure \ref{fig:channels_mirror_nojvm} of Appendix \ref{sec:figures} for non-JvM-corrected and non-continuum-subtracted images. In Figures \ref{fig:isovelocities_deproj}, \ref{fig:allchannels}, and \ref{fig:allchannels_nojvm} of Appendix \ref{sec:figures}, these signals are further illustrated in channel maps of both HCN and \CtwoH{} emission. Crucially, although the HCN intensity signal from the CPD candidate is spatially localised, it is significant ($>5\sigma$) across multiple velocity channels and can therefore be robustly exploited for dynamical analysis. Some additional processing is required, however, as the signal is partially obscured by overlapping velocity components from the bright upper layers of the circumstellar disc along the line of sight.

The fact that the CPD signature is not entirely hidden by these upper layers can be explained by the expected vertical and radial offset between the CPD emission, assumed to originate near the disc midplane, and the emission surface of the circumstellar disc intersecting the same line of sight. This offset produces an apparent spectral shift between the two components due to differences in orbital velocity, allowing the CPD emission to remain visible in a subset of velocity channels (see Sect. \ref{subsec:vertical_cpd} for a discussion of the vertical location of the proposed CPD signals).

To isolate the CPD emission and minimise contamination from the circumstellar disc, we leverage our knowledge of its three-dimensional structure (Sect. \ref{sec:discminer}) and apply a folding technique in which spectra from the blueshifted side of the disc are subtracted from those on the redshifted side in the vicinity of the CPD. This procedure is illustrated in Figure \ref{fig:channels_mirror}, with the reconstructed CPD spectra shown in the bottom panel, extracted from a $7\times7$ pixel region (two beams per side) centred on the localised HCN signal. The brightest spectrum has a Gaussian amplitude of 5.9\,mJy\,beam$^{-1}$, corresponding to a $7\sigma$ detection referred to the noise level of 0.8\,mJy\,beam$^{-1}$ in the same folded region, and a centroid velocity of 2.29\,\kms{} measured relative to the systemic velocity of the circumstellar disc. Based on the 20\% brightest spectra, we find a composite FWHM of 0.94\,\kms{}, a median CPD orbital radius of $R = 88 \pm 7$\,au and an azimuth of $\phi = 46 \pm 3^\circ$ in the disc frame, corresponding to $R = 0\farcs{75} \pm 0\farcs{05}$ and $\rm{PA} = 350 \pm 3^\circ$ in projected sky coordinates\footnote{The quoted uncertainties correspond to the standard deviation of the projected locations among the selected spectra.} (or $\Delta\alpha = -0\farcs{13}$, $\Delta\delta =0\farcs{74}$), relative to the central star. 

This demonstrates a strong quantitative overlap between the proposed CPD and the kinematic planet P94 reported at $R=94\pm6$\,au, $\phi=50\pm3^\circ$ by \citet{izquierdo+2022, izquierdo+2023}. Such an overlap is further supported by the fact that the radial location of planets inferred from localised velocity perturbations, such as P94, may be overestimated by 5–10\%, since this method appears to be slightly more sensitive to the outer planet-driven spiral wake than to the inner one, as demonstrated by previous modelling efforts \citep{izquierdo+2021, bae+2025}.

\begin{figure*}
   \centering
   \includegraphics[width=1\textwidth]{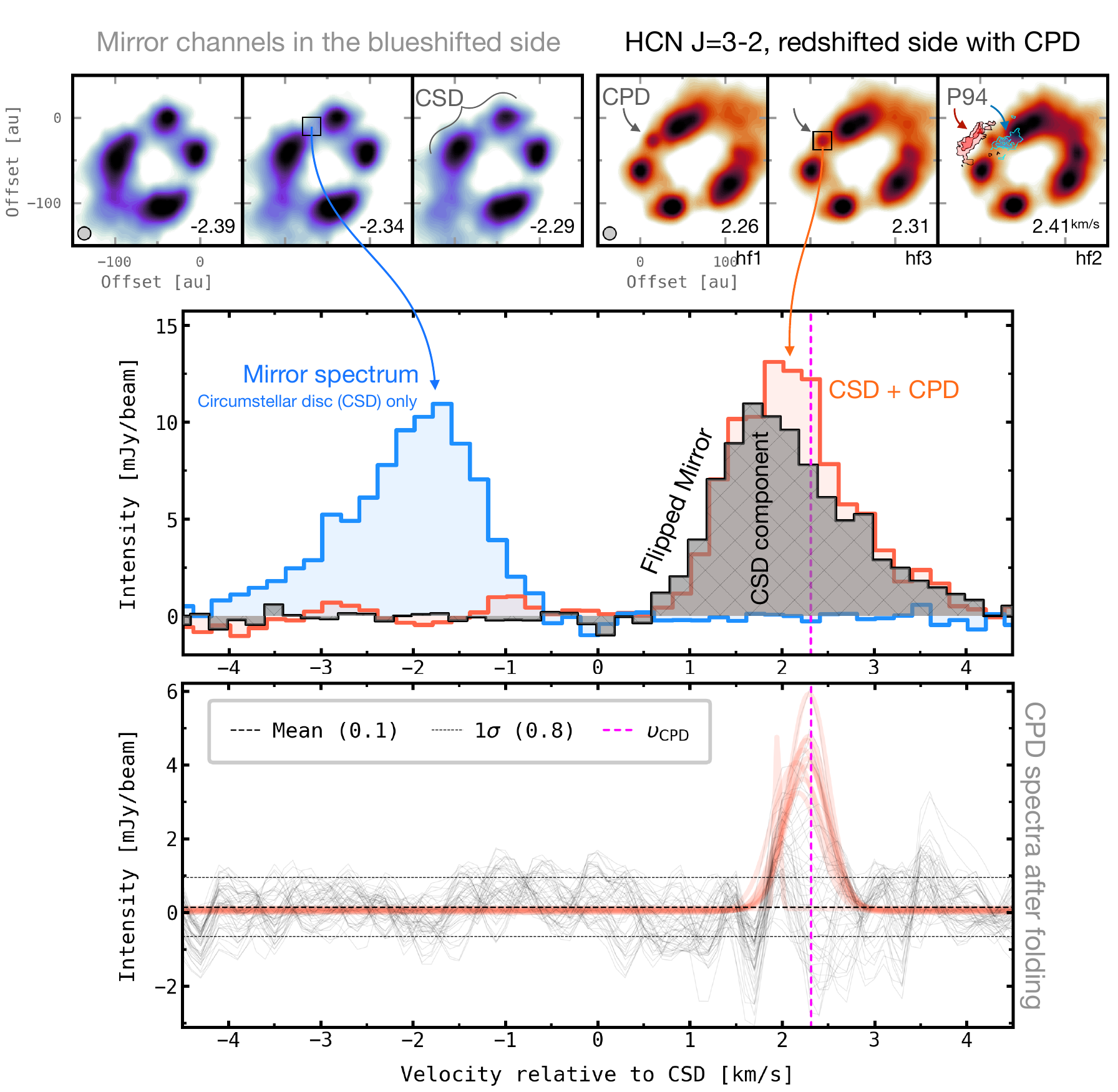} 
      \caption{Illustration of the folding procedure used to isolate the \hcnfull{} spectra from the localised signal around the kinematic planet P94, possibly tracing circumplanetary disc (CPD) emission. The blueshifted side of the circumstellar disc (CSD) is spectrally mirrored and subtracted from the redshifted side, where the planet resides, retaining the excess intensity contribution from the candidate CPD. The bottom panel shows the resulting folded spectra extracted from a $7\times7$ pixel box ($\sim\!2$ beams per side) centred on the proposed CPD location at $R = 0\farcs{75}$ and $\mathrm{PA} = 350^\circ$. Gaussian fits to the brightest 20\% of the spectra are overlaid as orange lines. The global peak profile has an amplitude of 5.9\,mJy\,beam$^{-1}$ and a centroid velocity of 2.29\,\kms{} marked by the vertical dashed line, measured with respect to the inferred systemic velocity of the CSD.
              }
         \label{fig:channels_mirror}
\end{figure*}

\subsection{Dynamical Planet Mass and CPD Properties}
\label{subsec:cpd}

Under the assumptions outlined later in this section, it is in principle possible to constrain some of the physical properties of the proposed CPD system, including the planet mass $M_p$, by noting that its line profile must be broadened by at least two components: the thermal motions of the gas, which contribute $\upsilon_{\rm th} = \sqrt{k_BT/m_{\rm HCN}}$ to the intrinsic width of the line, and the orbital rotation of material around the planet, which induces velocity shifts in the bulk line profile but can effectively broaden it when the observations are unresolved. These effects are further modulated by the CPD radius $R_{\rm CPD}$ and inclination $i_{\rm CPD}$, which, together with the CPD temperature $T$ also determine the observed line intensity per velocity channel.

To combine these ingredients while accounting for the finite resolution of the observations, we generate \discminer{} channel-map models of the reconstructed CPD emission. Since the CPD signal is spatially unresolved, we implemented a subpixeling routine in our models to capture variations in the intensity and velocity gradients of the CPD, as detailed in Appendix \ref{sec:subpixeling}, in addition to the standard gridding at the native pixel scale and subsequent convolution by the beam. The compactness of the signal is consistent with theoretical expectations from angular momentum conservation, which implies that material slowly accreting into the Hill sphere of a forming planet is truncated at approximately one-third of the Hill radius \citep{quillen+1998, ayliffe+2009}, defined as $R_{\rm Hill} = R_p(M_p / 3M_\star)^{1/3}$. For a 1\,\Mj{} planet at 88\,au, this yields a CPD radius of just $\sim\!1.5$\,au---well below the $\sim\!15$\,au beam size of the observations at the source distance. More recent 3D radiative hydrodynamic simulations have proposed slightly larger truncation scales, up to $\sim\!0.6\,R_{\rm Hill}$ \citep{szulagyi+2017}, yet still insufficient to be spatially resolved in the data\footnote{Even at this larger truncation radius, a prohibitively high companion mass of $15$\,\Mj{} would be required for the emission to become spatially resolved.}.

To reduce degeneracies among the key parameters of the CPD system that we aim to constrain ($M_p$, $R_{\rm CPD}$, and $T$), we assume that the CPD follows Keplerian rotation\footnote{As discussed in Sect. \ref{subsec:isothermal_adiabatic}, the strength and shape of the CPD’s rotational component depend not only on the planet’s mass but also, more generally, on its thermal history, which determines the relative contribution of pressure forces in supporting the gas around the planet. This complexity, however, lies beyond the scope of our simple parametric model, given the limitations imposed by the unresolved observations.}, is geometrically thin, and coplanar with the circumstellar disc (i.e. $i_{\rm CPD} = 46.7^\circ$). 
In addition, since the temperature of the CPD probed by HCN is currently unknown, we run nine separate models, each adopting a uniform gas temperature spanning values between $T = 25$\,K and $T = 800$\,K, with the thermal line broadening adjusted accordingly. The lowest temperature, 25\,K, corresponds to the rotational temperature inferred by \citet{bergner+2021} for the circumstellar disc, based on analysis of the HCN hyperfine structure, at the orbital radius of the CPD candidate. The higher-temperature models are intended to account for the planet's thermal/accretion heating, the CPD's internal viscous heating, and shock or compressional heating from infalling circumstellar material \citep{szulagyi+2016, szulagyi_mordasini+2017, cridland+2025}. Finally, we assume that the emission is optically thick, and thus the model peak intensity---prior to projection and convolution---is directly determined by the adopted temperature.

\setlength{\tabcolsep}{6pt}

\begin{table*}
\centering
{\renewcommand{\arraystretch}{1.5}
\caption{Median parameter values and 16th and 84th percentile uncertainties derived from the marginalized posterior distributions (final 1000 steps) of the CPD line profile models introduced in Sect. \ref{subsec:cpd}, for different gas temperatures. }
\label{table:cpd_parameters}

\begin{tabular}{lrrrrrrrrrc}
\hline
 & \textbf{25 K} & \textbf{50 K} & \textbf{75 K} & \textbf{100 K} & \textbf{150 K} & \textbf{200 K} & \textbf{400 K} & \textbf{600 K} & \textbf{800 K} & Unit \\
\hline
$M_p$ & $1.42_{-0.12}^{+0.10}$ & $1.08_{-0.11}^{+0.11}$ & $0.98_{-0.08}^{+0.07}$ & $1.29_{-0.23}^{+1.59}$ & $1.12_{-0.34}^{+0.27}$ & $0.65_{-0.22}^{+0.22}$ & $0.21_{-0.06}^{+0.10}$ & $0.09_{-0.06}^{+0.07}$ & $0.04_{-0.03}^{+0.05}$ & $M_\mathrm{Jup}$ \\
$R_\mathrm{CPD}$ & $7.45_{-0.09}^{+0.10}$ & $4.49_{-0.10}^{+0.06}$ & $3.33_{-0.07}^{+0.07}$ & $2.87_{-0.15}^{+0.09}$ & $2.10_{-0.05}^{+0.08}$ & $1.68_{-0.08}^{+0.05}$ & $1.02_{-0.04}^{+0.04}$ & $0.79_{-0.04}^{+0.05}$ & $0.64_{-0.05}^{+0.05}$ & au \\
$v_\mathrm{CPD}$ & $2.26_{-0.01}^{+0.01}$ & $2.26_{-0.04}^{+0.02}$ & $2.26_{-0.02}^{+0.01}$ & $2.31_{-0.03}^{+0.19}$ & $2.32_{-0.03}^{+0.03}$ & $2.31_{-0.02}^{+0.02}$ & $2.31_{-0.02}^{+0.02}$ & $2.35_{-0.02}^{+0.02}$ & $2.38_{-0.02}^{+0.02}$ & \kms{} \\

\bottomrule 

\end{tabular}
\justifying

{\noindent \textbf{Note.} The CPD centroid velocity, $v_\mathrm{CPD}$, is reported relative to the systemic velocity of the circumstellar disc, 5.79\,\kms{}.
  }
}
\end{table*}

As a result, each model run includes six free parameters: planet mass $M_p$, central offsets $x_p$ and $y_p$, centroid velocity $\upsilon_{\rm CPD}$ relative to the systemic velocity of the circumstellar disc, CPD position angle $\mathrm{PA}_{\rm CPD}$, and CPD radius $R_{\rm CPD}$. After an initial burn-in phase of 5000 steps, we find that the MCMC chains from all nine models have stabilized. The models were then run for an additional 15,000 steps to map the posterior distributions. Table \ref{table:cpd_parameters} summarises the median planet masses, CPD sizes, and systemic velocities, along with their associated uncertainties derived from the final 1000 steps of the parameter chains. Figure \ref{fig:cpdchannels} in Appendix \ref{sec:figures} illustrates the reconstructed CPD intensity channels together with the corresponding model channel maps for the different adopted temperatures.

\begin{figure}
   \centering
   \includegraphics[width=0.5\textwidth]{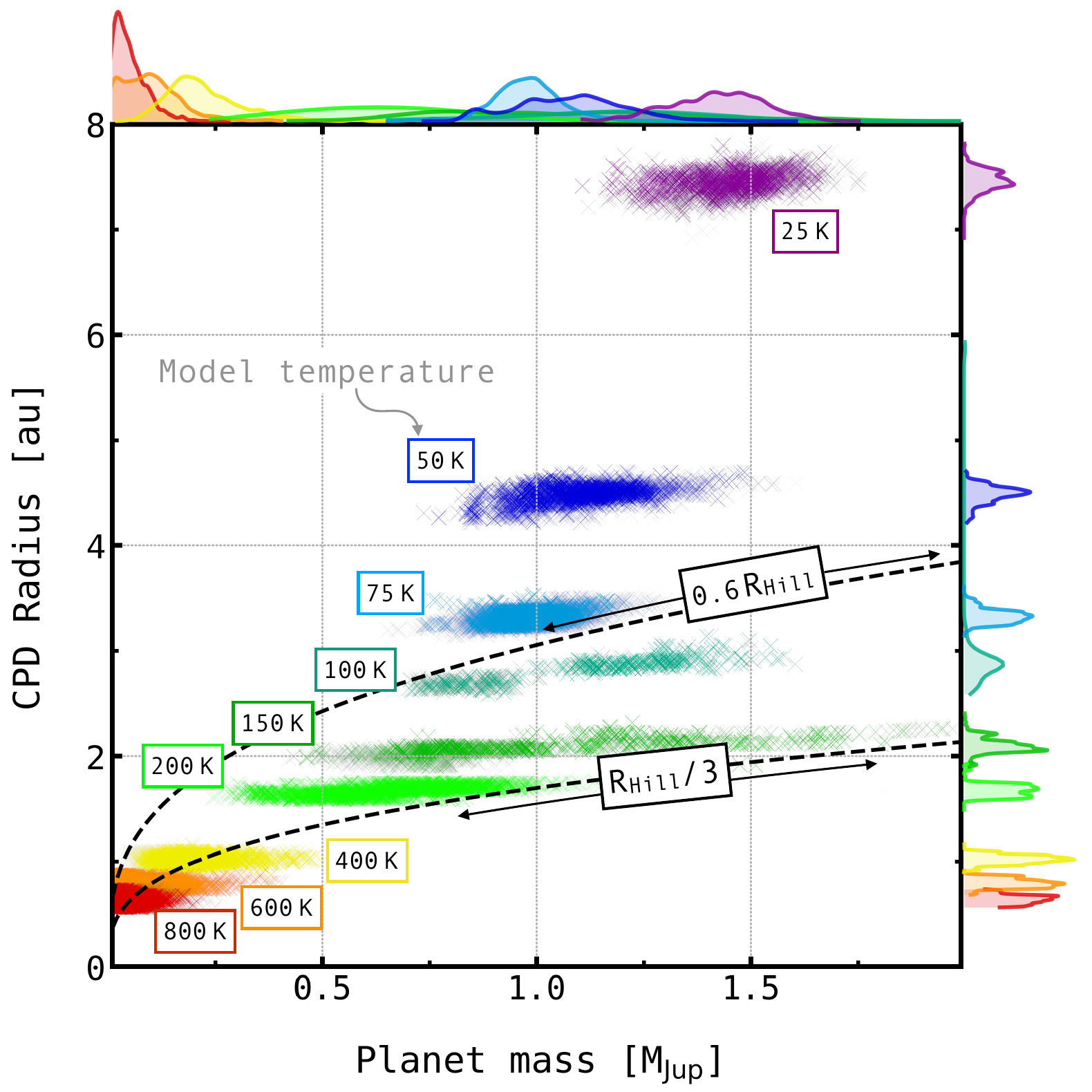} 
      \caption{CPD radii and planet masses from the posterior distributions derived from our channel-map models of the localised HCN signal around the planet candidate P94. Model temperatures above 150\,K yield CPD sizes consistent with theoretical expectations in which the CPD radius corresponds to roughly one-third of the planet’s Hill radius.
              }
         \label{fig:posteriors}
\end{figure}

The highest possible planet masses and CPD sizes are set by the coldest models at $T=25$\,K, which yield planet masses consistently below 1.8\,\Mj{} and CPD radii no larger than 8\,au. These two properties are compared across all models in Figure \ref{fig:posteriors}, which shows the posterior distributions relating CPD radius and inferred planet mass for each assumed temperature. Notably, the modelled planet mass range is broadly consistent---within a factor of two---with the previous estimate of 1\,\Mj{} derived from analyses of surface density and velocity perturbations at the same location in the circumstellar disc \citep{teague+2018a, izquierdo+2022}. 

More generally, we highlight that models with CPD gas temperatures of $T > 150$\,K, planet masses $M_{\rm p} < 1.0$\,\Mj{}, and CPD radii $R_{\rm CPD} < 2$\,au are most consistent with theoretical expectations in which the CPD outer boundary corresponds to approximately one‑third of the planet’s Hill radius. Considering the possibility of larger truncation scales $\sim\!0.6R_{\rm Hill}$ suggested by 3D radiative hydrodynamical simulations \citep[e.g.][]{szulagyi+2017}, these limits are relaxed by roughly 50\%, yielding $T > 100$\,K, $R_{\rm CPD} < 3$\,au, and $M_{\rm p} < 1.5$\,\Mj{}. At the high end of the adopted temperature range, our models best reproduce the data for $T < 400$\,K under our simplified assumptions of a uniform CPD temperature and Keplerian rotation. As illustrated in Fig.~\ref{fig:cpdchannels}, model temperatures above this threshold tend to flatten the line profile and overestimate the flux in the line wings. Nevertheless, the generally high temperatures favored by our models are in excellent agreement with a substantial increase in HCN abundance predicted by thermochemical models, where chemical reactions leading to HCN formation can overcome their activation barrier at a few hundred kelvin \citep[see][and Sect. \ref{subsec:chemistry} for a discussion]{doty+2002}, potentially explaining the origin of the detected signal.

Furthermore, the relatively low planet masses inferred from our analysis are consistent with the absence of hydrogen recombination lines (commonly used as tracers of ongoing accretion) in SPHERE/ZIMPOL data \citep{huelamo+2022} and the lack of significant infrared emission in JWST/NIRCam observations of \hd{} planet candidates, including P94 \citep{uyama+2025}. Indeed, theory indicates that extinction from the circumstellar disc and CPD may strongly attenuate both intrinsic luminosity and recombination-line emission, such that only planets with masses on the order of 2\,\Mj{} or higher can produce thermal and accretion signatures detectable with current optical and infrared facilities \citep{sanchis+2020, szulagyi+2020}. As such, anomalies in the surrounding chemistry, like the ones presented here, remain among the most powerful methods for directly investigating the properties of deeply embedded planets and their vicinities.

On the other hand, we note that the model CPD sizes are all consistently smaller than the width of the millimeter dust gap D85 \citep[$\sim$16\,au wide;][]{huang+2018_incl} in which the CPD resides, consistent with planet-disc interaction simulations predicting that millimeter gap widths can extend up to 10$R_{\rm Hill}$ \citep[e.g.][]{rosotti+2016}. This suggests that the abundance of millimeter‑sized grains in the circumplanetary region may be reduced owing to the limited inflow of particles aerodynamically trapped by pressure maxima at the gap walls, likely by the influence of the same parent planet P94  \citep{izquierdo+2023}. As a consequence, the associated millimeter flux is suppressed, reducing the likelihood of direct detection in the millimeter continuum as predicted by \citet{chrenko+2025}. In contrast, we note that this radial section of the disc is not fully depleted in gas, with multiple species still emitting prominently \citep[e.g., CO, HCO$^+$, H$_2$CO, CS,][]{law+2021_maps3}. This suggests that gaseous material can continue to flow through the gap either along the midplane or from the upper disc layers and accrete onto the CPD, potentially further enriching its chemical composition.

We emphasise that the planet masses and CPD radii derived from our models should be regarded as initial estimates rather than firm constraints. Our simplified modelling neglects additional sources of line broadening such as hydrodynamical turbulence, inflow of material onto the CPD, hyperfine blending, and opacity broadening, which would generally act to reduce the inferred planet masses if included. Conversely, as discussed in Sect. \ref{subsec:isothermal_adiabatic}, the omission of pressure support within the CPD leads to an underestimation of the planet mass. Future thermochemical and hydrodynamic models constrained by the observed features will improve the accuracy of the inferred CPD properties.

\section{Discussion} \label{sec:discussion}

\subsection{Origin of the Observed CPD Chemical Signatures} \label{subsec:chemistry}

The localised HCN and \CtwoH{} intensity signatures that we attribute to circumplanetary material around P94 may provide compelling evidence for an evolved chemical environment, sustained by high C/O ratios that favor carbon‑rich chemistry, together with enhanced UV flux \citep{cleeves+2015, bosman+2021} potentially driven by ongoing accretion onto the protoplanet \citep{aoyama+2018, aoyama+2020}. This has important implications for the composition of the nascent planet-satellite system as it could ultimately trigger the gas‑phase formation of more complex organics within the planet‑forming zone \citep{calahan+2023}.

Along with enhanced exposure to a planet‑driven UV field, we may also be witnessing the effects of local planetary heating, enabling reactions between H\textsubscript{2} and small organic radicals such as CN and C\textsubscript{2}. These reactions can overcome their activation barriers at temperatures of a few hundred kelvin---consistent with a subset of our CPD models---producing HCN and \CtwoH{} via CN+H\textsubscript{2}$\rightarrow$HCN+H and C\textsubscript{2}+H\textsubscript{2}$\rightarrow$\CtwoH{}+H \citep{pitts+1982, he+1998, doty+2002, ju+2006}. Such a mechanism can lead to enhancements in HCN abundances of up to two orders of magnitude at temperatures above $\sim\!200$\,K \citep[e.g.][as measured in high-mass hot cores]{boonman+2001}. Relatedly, as demonstrated by \citet{cleeves+2015}, even at relatively low temperatures just above $T\!\sim\!44$\,K, planetary heating from a Jupiter-mass planet can locally trigger the thermal desorption of HCN that would otherwise remain frozen out in the cold disc midplane. This process can significantly enhance the column density of the tracer in the planet’s vicinity and has therefore been regarded as a promising indicator of planet presence.

The temperatures required to activate either of these chemical pathways in the proposed CPD of P94 are consistent with our models that best match theoretical expectations for CPD sizes, which yield temperatures significantly higher ($T > 100$\,K) than those predicted for the circumstellar disc at the same location \citep[$T \sim 25$\,K;][]{law+2021_maps4}. This indicates that localised heating is indeed active and potentially influencing not only the CPD's dynamical structure (see Sect. \ref{subsec:isothermal_adiabatic}) but also the chemistry and observability of its molecular reservoir.

A similar conclusion was reached by \citet{bae+2022} for a candidate CPD embedded in the outskirts of the disc around the T Tauri star AS\,209, where CPD temperatures of at least 35\,K (exceeding the 22\,K of the background disc) were required to explain the observed localised $^{13}$CO emission. Despite differences in stellar type, evolutionary stage of the host stars, and orbital separation, this shared characteristic suggests that heating processes within CPDs may enhance the likelihood of detecting planetary seedbeds on wide orbits where the temperature contrast between the CPD and the circumstellar disc is expected to be substantial. \citet{chrenko+2025} explored an analogous scenario using radiative-hydrodynamic simulations of warm CO bubbles around a Jupiter‑mass planet at 120\,au from a solar-mass star, far enough for the bubble to stand out against the cold circumstellar disc background. Despite the use of a different tracer, the combined contributions of the planet-induced bubble and the circumstellar disc to the total line intensity appear strikingly similar to the signatures observed around P94 in \hd{}.

\subsection{Vertical Location of the CPD Emission} \label{subsec:vertical_cpd}

Another piece of evidence supporting the planetary origin of the localised signals is that they appear to arise from the disc midplane, where planet formation is most likely to occur as a result of the enhanced concentration of solids driven by hydrodynamical and gravitational instabilities (e.g., \citealt{youdin+2002}; \citealt{bai+2010}). This is evidenced by the fact that, at the proposed orbital separation of the planet, the estimated systemic velocity of the candidate CPD matches the expected Keplerian velocity at the midplane, indicating that the CPD emission originates below the elevated HCN layer of the circumstellar disc.

This is illustrated in Figure \ref{fig:isovelocities_deproj} (Appendix \ref{sec:figures}), where the localised intensity signal intersects the Keplerian isovelocity contour corresponding to the CPD systemic velocity only when the emission is deprojected assuming a midplane origin. Conversely, a consistent deprojection of the circumstellar disc emission is achieved only when adopting the appropriate emission surface height of $z/r\!\sim\!0.2$, as derived from \discminer{} or \disksurf{} modelling in Sect. \ref{subsec:height}.

\subsection{Circumplanetary Disc or Envelope? Implications from the Observed Rotational Velocity} \label{subsec:isothermal_adiabatic}

Hydrodynamical simulations of circumplanetary material carried out by \citet{fung+2019} indicate that the degree of rotational support within the CPD generally evolves over the planet’s thermal history, with a response that may strongly depend on the planet-to-star mass ratio. Specifically, the authors demonstrate that in the initial envelope stage, when the circumplanetary gas behaves adiabatically on timescales shorter than the cooling time, the rotational component is negligible, as the gas is largely supported by a steep radial pressure gradient. Over timescales longer than the cooling time, as the system evolves toward an isothermal state, the pressure support weakens and the rotational component increases, gradually flattening the circumplanetary material into a disclike structure. In this state, the rotation approaches Keplerian values in the inner regions but remains consistently sub-Keplerian in the outer regions owing to residual pressure support \citep[see also][for additional discussions]{lee+2015, coleman+2017, lega+2024}.

Assuming an early adiabatic scenario in which thermal broadening dominates the observed line width of the proposed CPD, we find that a temperature of $1150$\,K would be required to reproduce the composite FWHM of 0.94\,\kms{} exhibited by the HCN signal, following  $\upsilon_{\rm fwhm} \equiv 2.355\sqrt{k_BT/m_{\rm HCN}}$. Although this simplified estimate does not account for the finite resolution of the observations, it is consistent with our finding that for gas temperatures $T>800$\,K the rotational contribution of the CPD becomes negligible (as $M_p\rightarrow0$; see Sect. \ref{subsec:cpd}). This effect has far‑reaching implications for the inference of planet masses based on circumplanetary line emission, as early-stage, adiabatically heated configurations would largely suppress the observable kinematic imprint of the planet on the surrounding gas velocities \citep[see][for a discussion]{szulagyi+2017}.

On the other hand, depending on whether the planet formed through a cold‑start core accretion or hot‑start gravitational instability pathway, its effective temperature, and consequently its thermal feedback onto the CPD, may differ significantly over time, with direct consequences on their chemical observability. For example, spectral and photometric models of a 1\,Myr old Jupiter‑mass planet predict effective temperatures of  $\sim\!500$\,K for a cold‑start formation and $\sim\!800$\,K for a hot‑start pathway \citep{spiegel+2012}, implying that the observable chemical anomalies discussed in Sect. \ref{subsec:chemistry} would respond differently under each scenario. We therefore anticipate that future independent estimates of the CPD temperature, derived from multi‑wavelength studies and coupled with thermochemical models tailored to this system, will place strong constraints not only on the CPD’s evolutionary state but also on the planet’s formation pathway.

\section{Conclusions} \label{sec:conclusions}

We report the detection of localised emission from simple organic molecules, HCN and \CtwoH{}, spatially coincident with the Jupiter-mass planet P94---previously proposed to explain CO kinematic perturbations in the disc of \hd{}---and thus possibly arising from a circumplanetary disc (CPD) structure. Although the signal is partially obscured by the upper layers of the circumstellar disc, a folding analysis comparing the redshifted and blueshifted sides of the system enables an approximate reconstruction of the candidate CPD spectrum in HCN with a significance of $7\sigma$. The detected feature is centred at an orbital radius of $R = 88 \pm 7$\,au and an azimuth of $\phi = 46 \pm 3^\circ$ in the disc frame (or $R = 0\farcs{75} \pm 0\farcs{05}$ and $\rm{PA} = 350 \pm 3^\circ$ in projected sky coordinates).

From a first-order dynamical analysis of the CPD spectrum in HCN, we derive an upper limit on the planet mass of 1.8\,\Mj{} consistent with previous estimates from kinematics, and a CPD radius not exceeding 8\,au, under the assumption that the CPD is Keplerian and coplanar with the circumstellar disc. By comparing the CPD sizes obtained from our models with theoretical expectations where the CPD radius is roughly one-third of the planet's Hill radius, we favour CPD gas temperatures $T > 150$\,K, planet masses $M_{\rm p} < 1.0$\,\Mj{}, and CPD radii $R_{\rm CPD} < 2$\,au.

Our analysis provides the first dynamical constraints on the properties of a forming planet and direct evidence for the presence of organic molecules in its vicinity, potentially sustained by high temperatures and UV irradiation driven by ongoing accretion onto the protoplanet. This establishes chemical anomalies as a powerful avenue for detecting circumplanetary regions and constraining the properties of embedded planets, and a promising alternative in cases where thermal infrared emission and accretion lines are weak due to (i) low planet masses, provided sufficient planetary heating can overcome molecular activation barriers and desorption temperatures, or (ii) strong extinction from circumstellar and circumplanetary material.

\section*{Acknowledgments}

We thank the anonymous referee for their constructive comments and suggestions, which improved the quality of this work. 
Support for A.F.I was provided by NASA through the NASA Hubble Fellowship grant No. HST-HF2-51532.001-A awarded by the Space Telescope Science Institute, which is operated by the Association of Universities for Research in Astronomy, Inc., for NASA, under contract NAS5-26555.
E.F.v.D. acknowledges support from the ERC grant 101019751 MOLDISK.
M.B. and J.S. have received funding from the European Research Council (ERC) under the European Union’s Horizon 2020 research and innovation programme (PROTOPLANETS, grant agreement No. 101002188). This Letter makes use of the following ALMA data: ADS/JAO.ALMA\#2018.1.01055.L. ALMA is a partnership of ESO (representing its member states), NSF (USA) and NINS (Japan), together with NRC (Canada), NSTC and ASIAA (Taiwan), and KASI (Republic of Korea), in cooperation with the Republic of Chile. The Joint ALMA Observatory is operated by ESO, AUI/NRAO and NAOJ. The National Radio Astronomy Observatory is a facility of the National Science Foundation operated under cooperative agreement by Associated Universities, Inc.

\software{\textsc{astropy} \citep{Astropy_2022}, 
            \textsc{casa} \citep{casa}, 
            \textsc{cmasher} \citep{cmasher+2020}, 
            \discminer{} \citep{izquierdo+2021},
            \textsc{disksurf} \citep{disksurf}, 
            \textsc{emcee} \citep{foreman+2013}, 
            \textsc{matplotlib} \citep{Hunter_mpl}, 
            \textsc{numpy} \citep{harris_np}, 
            \textsc{scikit-image} \citep{scikit-image}, 
            \textsc{scipy} \citep{Virtanen_scipy}, 
            \textsc{radio-beam} \citep{radio-beam}, 
            \textsc{spectral-cube} \citep{spectral-cube}}

\vspace{2cm}

\bibliography{references}

\appendix

\section{Modelling of unresolved velocity gradients through subpixeling} \label{sec:subpixeling}

Because the CPD emission is localised, it is essential for the model to accurately capture variations in the assumed intensity, line width, and velocity parameterizations at unresolved scales before convolution. To achieve this, we implemented a subpixeling routine in \discminer{}, which enables a more accurate estimation of line profile properties on a pixel‑by‑pixel basis. Unlike the standard approach, at each step of the parameter search the peak intensity, width, and centroid velocity of the model lines are evaluated locally at the subpixel level. These values are then used to generate as many independent Gaussian line profiles as the number of subpixels, which are subsequently averaged to obtain the image pixel intensity for each velocity channel.

To resolve the CPD's radial extent by at least three elements, we adopt a grid of 11 subpixels per image pixel for models with temperatures $\leq\!200$\,K and 21 subpixels for those with temperatures $>\!200$\,K, corresponding to effective resolutions of 0.4\,au and 0.2\,au, respectively. We find that incorporating subpixeling can alter the amplitude and width of the resulting line profiles by up to $\pm20\%$ compared to the standard (non-subpixel) modelling.

\section{Supporting Figures}
\label{sec:figures}

\begin{figure*}
   \centering
   \includegraphics[width=1\textwidth]{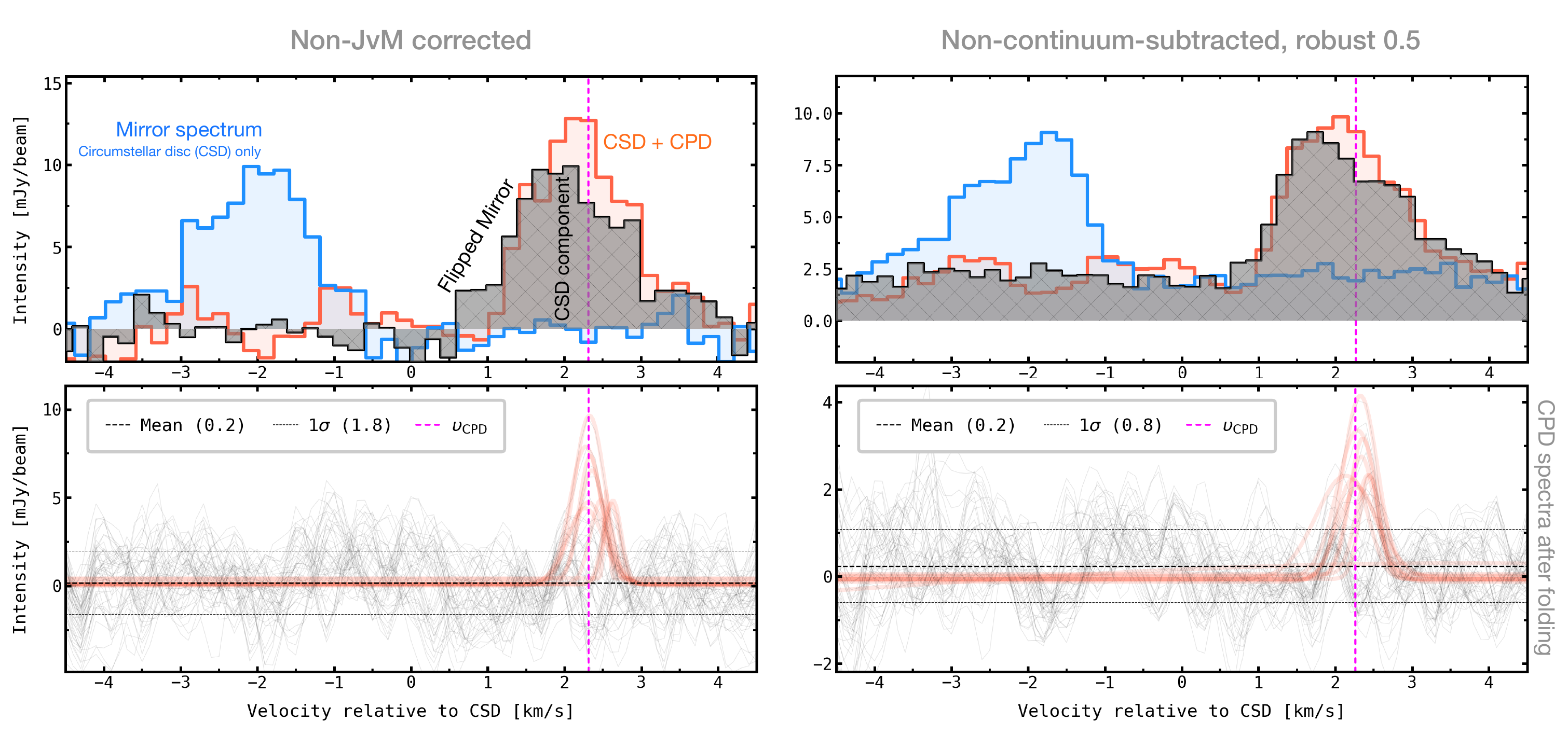} 
      \caption{Same as Figure \ref{fig:channels_mirror}, but for JvM-uncorrected and non-continuum-subtracted data cubes.
              }
         \label{fig:channels_mirror_nojvm}
\end{figure*}

\begin{figure*}
   \centering
   \includegraphics[width=1\textwidth]{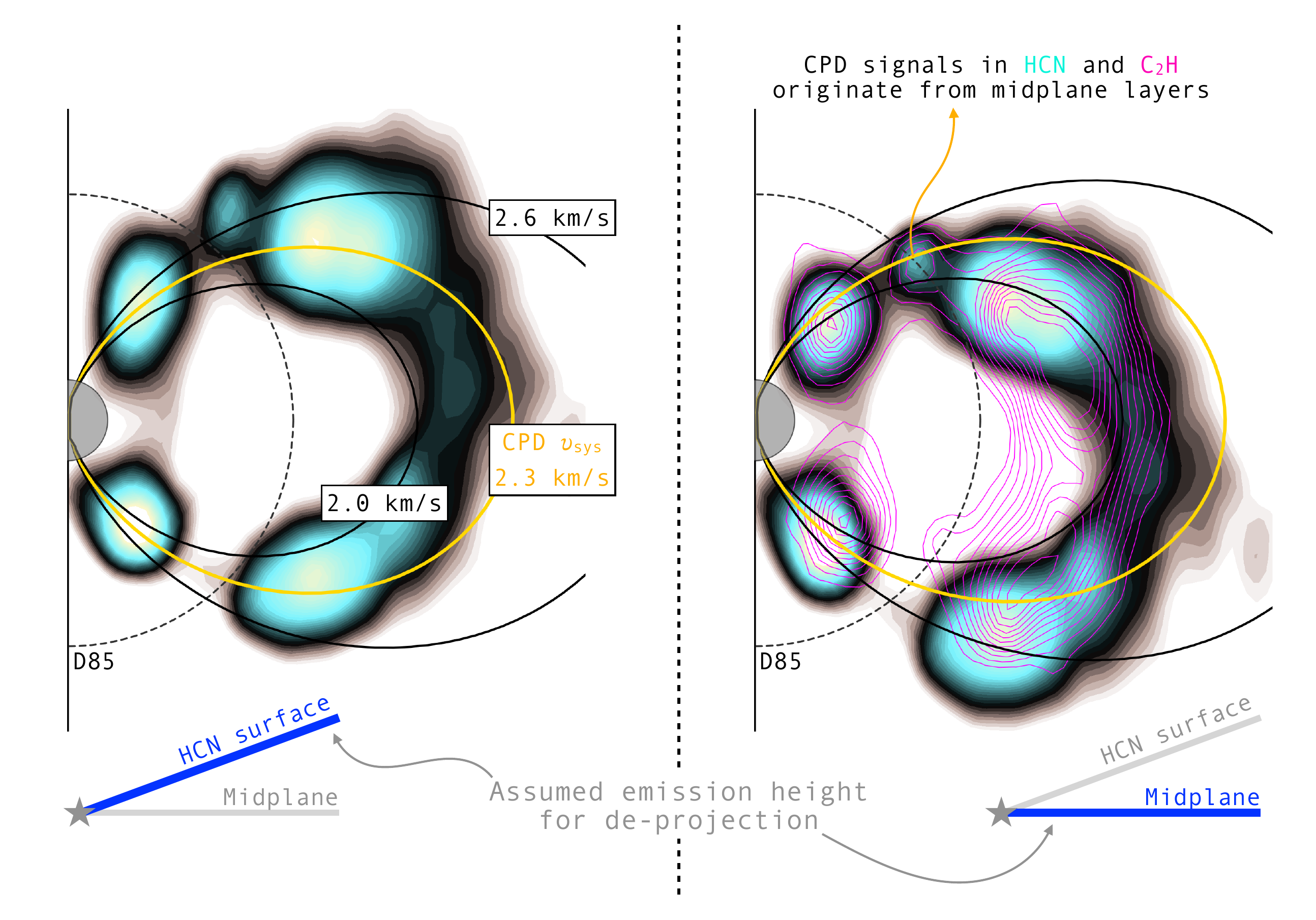} 
      \caption{Deprojected view of the disc of \hd{} as traced by HCN $J=3\rightarrow2$ emission in the 2.3\,\kms{} velocity channel relative to the circumstellar disc’s $\upsilon_{\rm LSRK}$, assuming elevated (left panel) and midplane (right panel) emission heights. Overlaid are Keplerian isovelocity curves at values around the inferred CPD centroid velocity ($\sim\!2.3$\,\kms{}; see Sect. \ref{subsec:cpd} for details). The observed CPD emission is consistent with a midplane origin, as indicated by its alignment with the isovelocity curve (yellow line) at the CPD centroid velocity when a $z/r = 0$ is adopted for the channel‑map deprojection. The right panel also includes C$_2$H emission (in purple contours) at the same velocity channel, with the CPD signal overlapping that seen in HCN.
              }
         \label{fig:isovelocities_deproj}
\end{figure*}

\begin{figure*}
   \centering
   \includegraphics[width=0.97\textwidth]{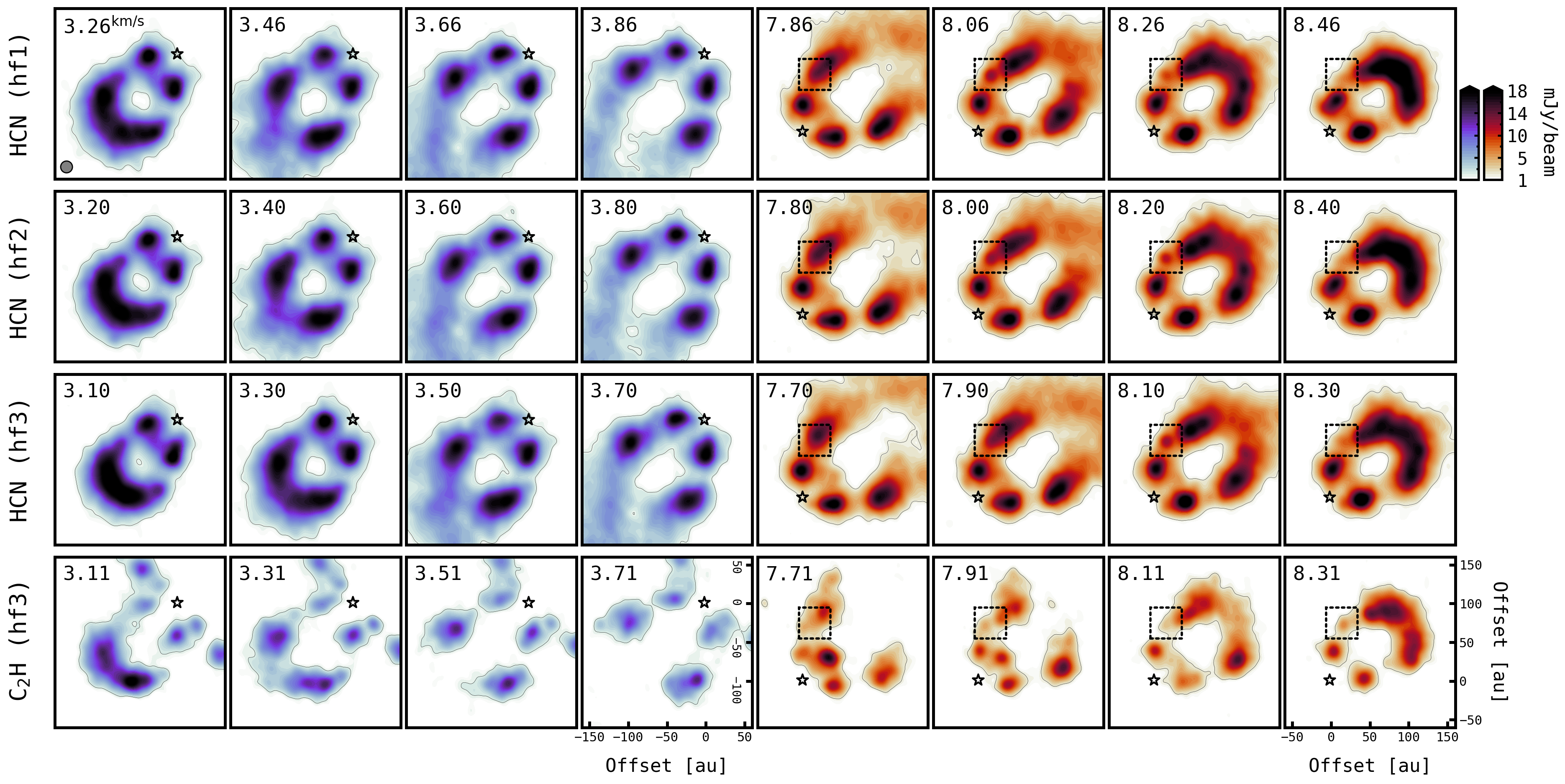} 
      \caption{Selected intensity maps for \hcnfull{} and \CtwoHfull{} $F=3\rightarrow2$ lines, shown at mirrored (blue) and around (red) velocity channels where a localised signal overlaps with the kinematic planet candidate P94 first proposed by \citep{izquierdo+2022, izquierdo+2023}, providing unprecedented evidence of circumplanetary material orbiting a still-forming planet (see Sect. \ref{subsec:cpd}). A guiding dashed box is centred at the inferred location of the signal, at a projected separation of $R = 0\farcs{75}$ from the image centre and a position angle of $\mathrm{PA} = 350^\circ$, measured from north through east. Although the hf1, hf2, and hf3 cubes for the HCN intensity channels correspond to datasets imaged at the central frequencies of different hyperfine components in the $J=3\rightarrow2$ group, the bulk of the emission in all cases originates from the dominant $F=4\rightarrow3$ line, blended with the $F=2\rightarrow1$ and $F=3\rightarrow2$ components. Contours enclose emission above five times the rms noise level of 0.43\,mJy\,beam$^{-1}$ for HCN and 0.53\,mJy\,beam$^{-1}$ for C$_2$H.
              }
         \label{fig:allchannels}
\end{figure*}

\begin{figure*}
   \centering
   \includegraphics[width=0.97\textwidth]{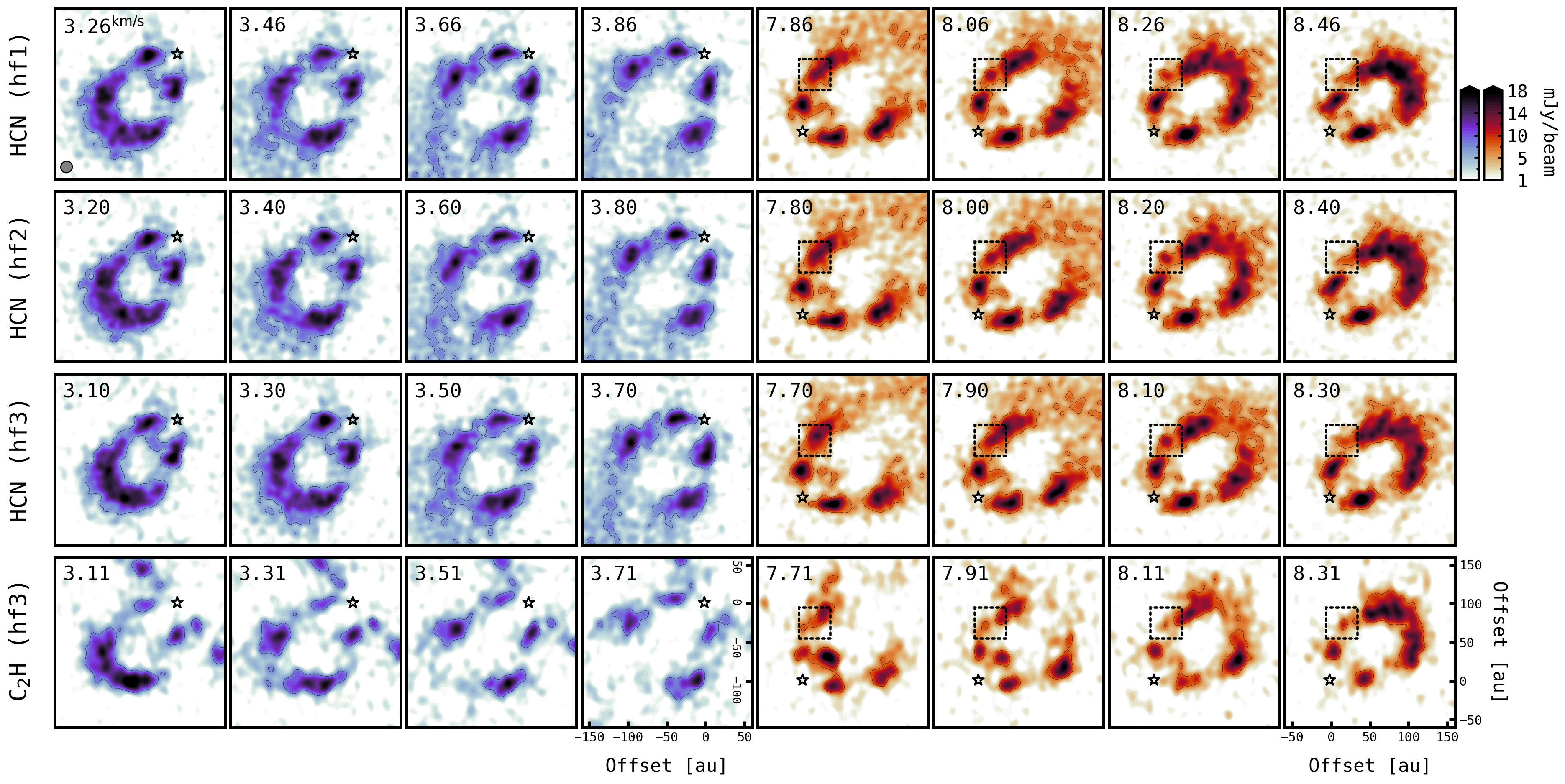} 
      \caption{Same as Figure \ref{fig:allchannels}, but for JvM-uncorrected images. The rms noise levels are 1.38\,mJy\,beam$^{-1}$ for HCN and 1.61\,mJy\,beam$^{-1}$ for C$_2$H in these data cubes.
}              
         \label{fig:allchannels_nojvm}
\end{figure*}

\begin{figure*}
   \centering
   \includegraphics[width=1\textwidth]{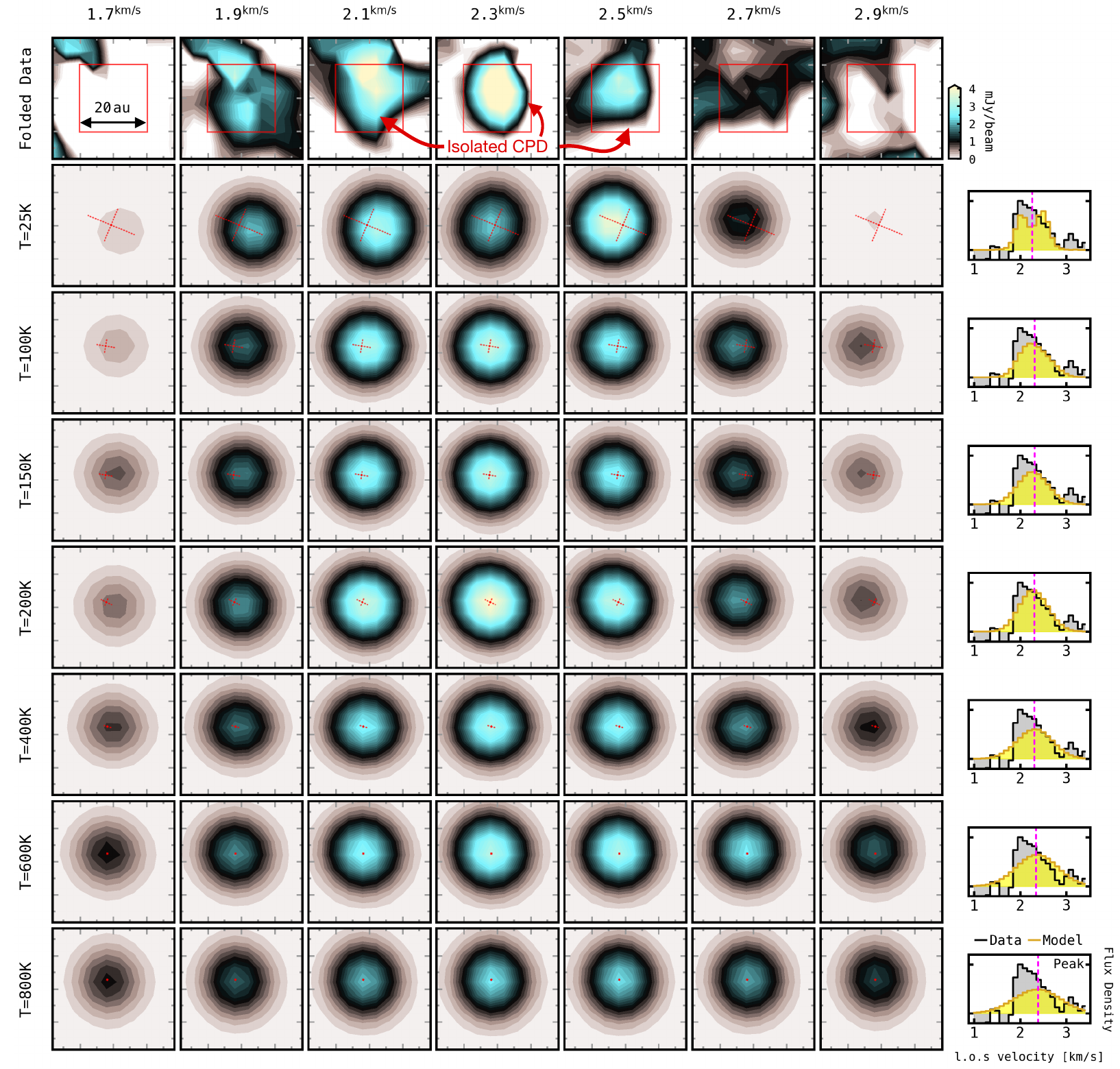} 
      \caption{Folded HCN channel maps centred on the location of the P94 planet, with a field of view of two beam sizes, illustrating the isolated CPD signal (top row) and the corresponding best-fit models introduced in Sect. \ref{subsec:cpd} for selected gas temperatures. The velocity channels indicated at the top are relative to the systemic velocity of the circumstellar disc, 5.79\,\kms{}. The dotted lines overlaid on the model channels indicate the location and orientation of the projected CPD major and minor axes derived from the fit. The rightmost column shows a comparison between the flux density from the data (black) and the models (yellow), extracted from the red box shown in the top row and normalised to the peak data value of 4.66\,mJy.
              }
         \label{fig:cpdchannels}
\end{figure*}

Here we provide complementary figures that support the results presented in the main text. Figure \ref{fig:channels_mirror_nojvm} illustrates the isolated \hcnfull{} signal obtained from non-JvM-corrected and non-continuum-subtracted images, following the analysis described in Sect. \ref{subsec:signals}. Figure \ref{fig:isovelocities_deproj} presents a deprojected view of one of the HCN channel maps containing emission from the CPD candidate, overlaid with the \CtwoH{} signal, highlighting the close alignment of the emission with the Keplerian isovelocity curve of the channel when the CPD signal is assumed to originate near the disc midplane (see Sect. \ref{subsec:vertical_cpd} for discussion).
Figures \ref{fig:allchannels} and \ref{fig:allchannels_nojvm} show HCN and \CtwoH{} channel maps around the detected signals for the JvM-corrected and non-JvM-corrected images, respectively. Finally, Figure \ref{fig:cpdchannels} summarises the best-fit channel maps obtained in Sect. \ref{subsec:cpd} for the HCN emission from the candidate CPD at selected model temperatures, together with a comparison of the corresponding flux-density profiles.

\end{document}